\documentclass{emulateapj}
\usepackage{amsmath, natbib}
\begin{document}

\title{Probing Electron-Capture Supernovae: X-Ray Binaries in Starbursts}
\author{T. Linden$^{1 , 2}$, J. F. Sepinsky$^{3 , 2}$, V. Kalogera$^2$, and K. Belczynski$^4$}
\affil{$^1$ Department of Physics, University of California, Santa Cruz, 1156 High Street, Santa Cruz, CA, 95064, USA}
\affil{$^2$ Department of Physics and Astronomy, Northwestern University, 2145 Sheridan Road, Evanston, IL 60208, USA}
\affil{$^3$ Department of Physics and Electrical Engineering, The University of Scranton, Scranton, PA 18510, USA}
\affil{$^4$ Oppenheimer Fellow; Los Alamos National Laboratory, Los Alamos, NM 87545, USA}
\slugcomment{Received 2009 February 22; accepted 2009 May 11; published 2009 June 24}

\shortauthors{}

\begin{abstract}
We develop population models of high-mass X-ray binaries (HMXBs) formed after bursts of star formation and we investigate the effect of electron-capture supernovae (ECS) of massive ONeMg white dwarfs and the hypothesis that ECS events are associated with typically low supernova kicks imparted to the nascent neutron stars. We identify an interesting ECS {\em bump} in the time evolution of HMXB numbers; this bump is caused by significantly increased production of wind-fed HMXBs 20--60\,Myr post starburst. The amplitude and age extent of the ECS bump depend on the strength of ECS kicks and the mass range of ECS progenitors. We also find that ECS-HMXBs form through a specific evolutionary channel that is expected to lead to binaries with Be donors in wide orbits. These characteristics, along with their sensitivity to ECS properties, provide us with an intriguing opportunity to probe ECS physics and progenitors through studies of starbursts of different ages. Specifically, the case of the Small Magellanic Cloud, with a significant observed population of Be~HMXBs and starburst activity 30--60\,Myr ago, arises as a promising laboratory for understanding the role of ECS in neutron star formation. 
\end{abstract}

\section{Introduction and Background}

Recently, electron-capture supernovae (ECS) of massive ONeMg white dwarfs have attracted renewed attention as an important mechanism of neutron star (NS) formation. ECS events probably operate at the low end of the NS progenitor mass range, when a strongly degenerate ONeMg core forms. If the progenitor exceeds a critical core mass of 1.38~M$_\odot$, high pressure and density lead to electron captures onto $^{24}$Mg and $^{20}$Ne. The resulting decrease in electron degeneracy pressure causes a supernova (SN) collapse before an iron core is ever formed \citep{1980PASJ...32..303M, 1984ApJ...277..791N, 1987ApJ...322..206N}. Numerical simulations indicate that ECS events are under-energetic compared to iron-core supernovae (ICC-SN\footnote{Here we use the term ICC to refer to both iron-core collapse supernovae, as well as direct collapse black holes (BHs) which form without a SN explosion (i.e, all non-ECS compact object formation events). We use ICC-SN to refer to only those iron-core collapse systems which result in an energetic supernova explosion}) by at least an order of magnitude, However they are, at the same time, relatively fast, neutrino-driven, delayed explosions. The short time available for the propagation of the shock is expected to prevent a strong build-up of core asymmetries, leading to typical kicks of $\lesssim 100$\,km\,s$^{-1}$,  lower than those in ICC-SN events \citep[][Janka 2007, private communication]{2006ApJ...644.1063D, 2006A&A...450..345K}.

The mass range of ECS progenitors is highly uncertain. While \citet{1984ApJ...277..791N, 1987ApJ...322..206N} predicted an initial mass range of 8-10~M$_\odot$ at solar metallicity, recent simulations suggest much stricter ECS boundaries for single stars. For example, \citet{2008ApJ...675..614P} find a single star mass range of 9-9.25~M$_\odot$, which accounts for only $\simeq$4\% of SNe for a given population. \citet{2007A&A...476..893S} finds a metallicity-dependent mass range which stays consistently narrow (a width of $\lesssim$1.5\,M$_\odot$), and could vanish altogether depending on the metallicity dependence of the stellar winds of evolved stars.

\citet{2004ApJ...612.1044P} proposed that binary evolution can result in a much wider ECS initial mass range of 8-11~M$_\odot$ in the following way. Stripping of the hydrogen-rich stellar envelope due to mass transfer and/or common envelope evolution aborts the second dredge-up phase, normally seen in single stars with their hydrogen envelope intact. Consequently the mass of the helium core is preserved, leading to relatively massive ONeMg cores that experience electron-capture collapse. They find that, in the closest binaries, even the first dredge up phase may be aborted, enabling ECS activity in the even lower mass range of 6-8~M$_\odot$.

Given the uncertainties involved, probing the ECS physical mechanism and its relevance to NS formation compared to ICC-SN events is a challenge. If ECS events are indeed associated with lower kicks \citep{2006ApJ...644.1063D, 2006A&A...450..345K}, then close binaries containing a NS are good candidate probes as their formation rates and properties significantly depend on SN kicks and mass loss. \citet{2002ApJ...574..364P} highlighted a sub-class of high-mass X-ray binaries (HMXBs) with low eccentricities that can be explained by low NS kicks. \citet{2004ApJ...612.1044P} and \citet{2004ESASP.552..185V, 2007AIPC..924..598V} postulated that the ECS mechanism could be responsible for these low kicks, although a very wide ECS progenitor mass range may be required to account for the observed HMXBs. 

In this {\em paper}, we present results that uncover an alternative probe for ECS events through studies of HMXB formation in starbursts of varying ages. We find that ECS NS formation associated with low kicks (lower than $\sim$50\,km\,s$^{-1}$) dramatically increases the formation efficiency of Be X-ray binaries in starburst ages of $\simeq20-60$\,Myr. In what follows, we discuss this effect in detail and analyze the reasons for this unique dependence, with the hope that observations of starbursts can be used to assess ECS kick velocities and progenitor mass ranges. We note that, in the Small Magellenic Cloud (SMC), dominated by $\simeq 40$\,Myr-old stellar populations, a systematic overabundance of Be XRBs has been observed \citep{2000A&A...359..573H, 2004ApJ...609..133M} potentially making the SMC a unique laboratory for ECS NS formation. 

\section{Simulation Code and Models}

For our modeling of HMXB formation and evolution in starbursts we employ a sophisticated population synthesis code, $StarTrack$, described in extensive detail in \citet{2008ApJS..174..223B}, which assumes that stellar dynamical interactions are not significant compared to the effects of binary evolution for the star formation conditions under consideration. We note that the parameter space of Monte Carlo population synthesis is very large, and thus a full exploration is not possible. Instead, we consider a default model (described in detail in \citet{2008ApJS..174..223B}) and we vary some binary evolution and ECS-related parameters with the purpose of highlighting their effects on the proposed ECS probe. Here we briefly summarize the main assumptions and parameters relevant to this study. 
 
We employ a delta function star formation episode at solar metallicity adopting: (1) a Salpeter (M$^{-2.35}$) initial mass function with primary masses above 4~M$_\odot$ and secondary masses above 0.15~M$_\odot$; (2) a flat mass ratio distribution; (3) a distribution of initial binary separations that is flat in the logarithm with an upper limit of 10$^5$ R$_\odot$ and a lower limit such that the primary star initially fills at most half of its Roche Lobe \citep{1983ARA&A..21..343A}, and (4) a thermal distribution for initial eccentricities \citep{1975MNRAS.173..729H}. We set the maximum NS mass at 2.5~M$_\odot$ and draw natal kicks for ``standard'' ICC-SN events from a single Maxwellian kick distribution with mean 265~km~s$^{-1}$ \citep{2005MNRAS.360..974H}. Kicks are potentially associated with BH formation as well (see \citet{2009ApJ...697.1057F} for the strongest evidence at present for a BH kick). For BHs formed through SNe explosions and subsequent fallback of material we multiply the normal Maxwellian kick by the fraction of the SN ejecta which is ultimately lost from the system. Usually BH formation at higher masses through direct collapse is assumed to be perfectly symmetric; here we adopt a small kick (10\% of the NS kick), for reasons discussed in detail in Section~\ref{disc}.

To obtain statistically significant results, we sample at least 10$^6$ initial binaries for each set of input parameters and we evolve the systems for 100\,Myr. However, we stress that the models and XRB numbers presented here are not normalized to match any specific observed system, since  we are purely interested in examining the {\em relative} XRB numbers as a function of starburst age. 
	
In order to compare our results with {\em Chandra} observations, we must determine the X-ray luminosity, $L_X$, based on the calculated mass-transfer rate for both wind-fed and Roche-lobe-overflow systems. We apply an estimated correction for the {\em Chandra} energy band as described and justified in \citet{ 2008ApJS..174..223B} (Section 9.1).  We analyze the HMXB population with $L_X$ in excess of $10^{32}$\,erg~\,s$^{-1}$, appropriate for observations of nearby star-forming galaxies. 
	
An important problem in the population synthesis of binary stars is the treatment of common envelope (CE) phases. In our simulations, CE events are treated using the usual energy formalism described by \citet{1984ApJ...277..355W}\footnote{In this treatment of CE events, the energy needed for CE ejection is removed from the binary orbital energy, often resulting in evolved stellar systems with very tight orbits.}, for stars which have established a clear core-envelope boundary. In this paper, we adopt a value for the CE efficiency of $\alpha_{CE}$ = 1. However, in our default models, we assume that a CE phase involving a donor star on the main sequence (MS) or hertzsprung gap (HG) leads to a binary merger \citep{2008ApJS..174..223B, 2000ARA&A..38..113T}. 

Following \citet{2000MNRAS.315..543H}, massive stars are assumed to explode in ECS events if the He core mass at the beginning of the asymptotic giant branch (AGB) is between 1.83-2.25 M$_\odot$ (see also \citet{2007arXiv0706.4096I}). This choice, in effect, selects specific initial mass ranges which depend on the binary evolution of the progenitors. However, given the uncertainties involved in selecting this mass range, we examine a number of He core mass ranges in what follows. We assume that ECS natal kicks follow a Maxwellian distribution with a smaller mean than ICC-SN kicks, thus we linearly scale down the ``standard'' ICC-SN Maxwellian distribution by varying factors. The mass of ECS-formed NS is assumed equal to 1.26\,M$_\odot$ \citep{1987ApJ...322..206N, 2004ApJ...612.1044P, 2007arXiv0706.4096I}, although we note that \citet{2006ApJ...644.1063D}, \citet{2006A&A...450..345K} and \citet{2008ApJ...675..614P} find slightly higher end masses for single star ECS events (1.338 - 1.4 M$_\odot$).

\section{The Role of Electron-Capture Supernovae in HMXB Production}

\begin{figure}
		\plotone{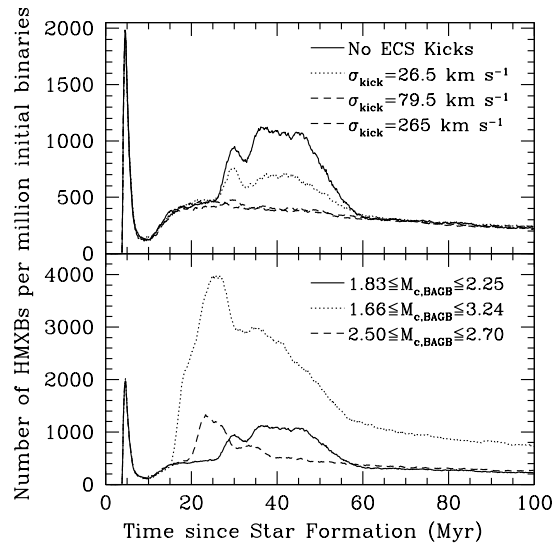}
		\caption{Number of HMXBs with L$_X$ $>$ 1 x 10$^{32}$\,erg\,s$^{-1}$ per $10^6$ initial binaries vs.\ starburst age for: (top) different Maxwellian sigma values for ECS natal kicks, and (bottom) different core mass ranges at the base of the AGB (BAGB) and no ECS kicks}
\label{ecsplot}
\end{figure}

Examination of our simulation results in terms of the number of HMXBs as a function of starburst age reveals an intriguing role of ECS NS formation: ECS events can dominate HMXB formation between 20 and 60\,Myr post-starburst, creating a clearly identified ``bump'' in the time evolution of HMXB numbers (see Figure~\ref{ecsplot}).  At any given metallicity, the width and relative height of this {\em ECS bump} is primarily dependent on two parameters: the typical ECS natal kick magnitude and the mass range of ECS progenitors. 

We first examine the effect of ECS kicks (Figure~\ref{ecsplot}, top panel). We typically find two significant bursts of HMXB activity: (1) a sharp, narrow peak at 5--10\,Myr post-starburst dominated by wind accretion from massive donors onto massive black holes formed in ``standard'' ICC events. As a result, this peak is, of course, insensitive to assumptions about ECS kicks; (2) a broader ECS bump (at 20--60\,Myr) dominated by wind accretion from lower mass donors onto NS. The sensitivity of this bump on ECS kicks is evident, as it effectively disappears and blends into the background HMXB population once typical ECS kicks significantly exceed $\sim50$\,km\,s$^{-1}$. 

As one might naturally expect, the ECS bump amplitude and age spread also depends on the AGB He core mass range for ECS events (Figure~\ref{ecsplot}, bottom panel). We find that more massive cores lead to earlier ECS bumps, as higher-mass cores select more massive stars that form NSs earlier. The effects of binary evolution change the mapping between He core masses and zero-age MS progenitor masses for ECS. The details of this mapping will depend on the specific stellar models adopted, but for the three mass ranges shown in Figure~\ref{ecsplot}  (bottom panel) and the $StarTrack$ models, the corresponding progenitor masses are found in the following ranges: 7.5--8.5\,M$_\odot$ and 9--11\,M$_\odot$ (solid), 11--14.5\,M$_\odot$ (dotted), and 9--9.5\,M$_\odot$ and 11.5--12.5\,M$_\odot$ (dashed). The change in the number of bright HMXBs for different ECS mass regimes by nearly a factor of 10 conveys the potential of this bump for constraining the details of the ECS mechanism and its dependence on the physical properties of the collapsing star.

Analysis of the evolutionary pathways of HMXBs reveals that CE evolution strongly governs the parameter space of bright HMXB progenitors. CE events leading to binary mergers greatly limit this parameter space by effectively imposing a minimum initial binary separation for avoiding CE mergers. Wider systems (with more evolved donors) may survive the CE event, acquiring tight post-CE orbits which contribute to HMXB production in two ways: (1) for CEs occurring prior to NS formation, tight orbits promote survival of the SN kick and (2) for CEs occurring after NS formation, tight orbits lead to stronger wind accretion and brighter HMXBs. 

\begin{figure}
		\plotone{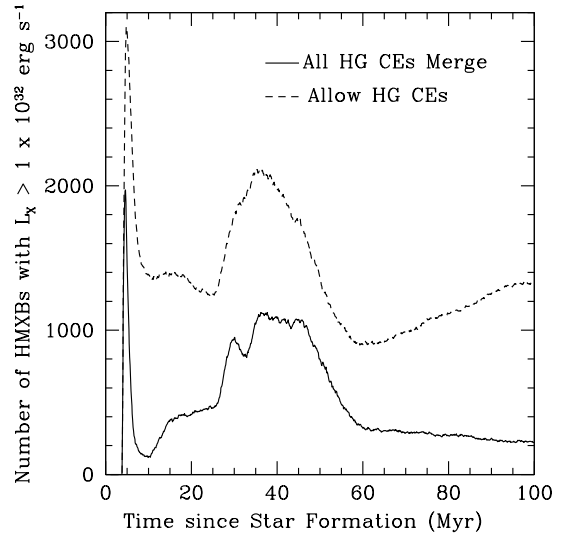}
		\caption{Number of HMXBs with L$_X$ $>$ 1 x 10$^{32}$\,erg\,s$^{-1}$ when CEs involving HG donors are always assumed to lead to binary mergers (solid curve) and when CEs with HG donors are treated with the usual \citet{1984ApJ...277..355W} energy formalism, which may lead to either mergers or tight post-CE binaries.}
		\label{ces}
\end{figure}

As already mentioned, in our default simulations we follow the findings of \citet{2000ARA&A..38..113T, 2004ApJ...601.1058I} and assume that CEs driven by MS or HG donors always lead to binary mergers. To examine the robustness of the ECS bump, we consider models where CEs driven by HG donors are also treated with the energy formalism. In this case more HMXBs form, but the presence of the ECS bump remains robust (Figure~\ref{ces}). Careful examination of the results indicates that the HMXB numbers populating both peaks (narrow and ECS) remain roughly constant, and the details of the CE treatment affect only the number of background (outside the peaks) systems throughout post-starburst evolution. 

We find that the resilience of both our BH-HMXB and ECS-HMXB peaks to CE events stems from their evolution through a particular channel and mass transfer history. Beginning with initial mass ratios close to unity, and relatively tight orbits, these systems go through mass transfer before BH/NS formation, but avoid a CE phase. Instead, they evolve through a mass ratio inversion which significantly expands their orbits through prolonged stable mass transfer. This pathway is dominant due to three main effects related to the initial binary properties, CE outcomes, and kick distributions of ECS and massive BH. 

The logarithmically flat distribution of initial orbital separations tends to favor tight systems. These undergo Roche-lobe overflow when the primary is moderately evolved (late on the MS, on the HG, or early in the core He burning [CHeB] phase). Given the mergers associated with CE events, we find that the majority of the surviving binaries have comparable masses (with accretor mass at least half the donor mass), ensuring stability of the mass transfer initiated by the original primary and avoiding a CE merger. Stable mass transfer soon reverses the binary mass ratio and the orbits begin to expand. This expansion continues until the hydrogen envelope of the original primary is exhausted, creating systems which are both evolved and relatively loosely bound (orbital separations of $\sim$500--2500\,R$_\odot$). Consequently, when the original primary explodes in a SN, binary survival is strongly favored if SN natal kicks are absent or low, which is the case for massive BH and for NS formed through ECS events. HMXBs in the ECS bump (Figure~\ref{ecsplot}) have relatively wide orbits and reach significant X-ray luminosities only when the donors (original secondaries) evolve off the MS onto the HG and CHeB phases where stellar winds are sufficiently strong. Lastly we note that, because of the comparable masses of the initial binary components, the post-MS evolution of the donor follows soon after the SN explosion of the primary, thus dictating the timing of the two identified peaks.  

Through this evolutionary sequence, ECS events (assumed to be associated with relatively low  SN kicks) lead predominantly to the formation of wide HMXBs with evolved donors that have reached the HG or the CHeB phase. The progenitors of these ECS-HMXBs have avoided a CE phase before the formation of the compact object and instead go through stable mass transfer from the original primary. Some systems can further evolve through a CE phase and form tightly bound HMXBs with He-rich donors, the relative contribution of which depends strongly on the assumed CE efficiency.

On the other hand, binaries that evolve through a CE phase before compact object formation end up with tight enough orbits to survive ICC-SN events with typical natal kicks in excess of $\sim100$\,km\,s$^{-1}$. These binaries tend to remain tight after the SN explosion and can become bright HMXBs even with unevolved, MS donors. These systems form a background of short period HMXBs through which the ECS bump can rise depending on the how favorable ECS conditions are (progenitor mass range and natal kicks). Since the creation of post-CE HMXBs is not dependent on the evolution of the donor, most MS-HMXBs form immediately after the SN of the primary ($<25$\,Myr) and their numbers exponentially decay thereafter due to subsequent Roche-lobe overflow by the MS donor and ultimately mergers. 

Lastly, we note that the formation of systems with wide orbital periods (greater than $\sim10$ days) is highly unlikely to result from ICC-SN activity involving large SN natal kicks (i.e. NS from non-ECS events and low-mass BH from ICC supernovae and fallback). Instead this population is dominated by ECS NS and BH formed through direct collapse, given the small kicks adopted for such events. However, as shown for ECS-HMXB in Figure~\ref{ecsplot}~(top), even these small kicks are enough to disrupt the widest of HMXBs.

\section{Discussion}
\label{disc}

We have shown that, if NS formation through ECS events is associated with natal kicks smaller than $\sim50$\,km\,s$^{-1}$, production of relatively bright wind-fed HMXBs (L$_X$ $>$ 1 x 10$^{32}$\,erg\,s$^{-1}$) is significantly favored between 20 and 60\,Myr after delta function star formation. We call this increased HMXB formation, which appears above a background HMXB population formed through standard ICC compact object formation, the ECS {\it bump}. The width in post-starburst age of this ECS bump and its amplitude relative to the rest of the ICC-HMXB population\footnote{including a narrow peak of HMXBs with massive BHs that appears earlier at $\sim5$\,Myr}  depend primarily on two ECS factors: the mass range of ECS progenitors and the typical magnitude of ECS natal kicks (Figure~\ref{ecsplot}). Moreover, we find that these ECS-HMXBs form through a specific evolutionary channel that avoids a CE phase before the SN and instead includes stable mass transfer from the primary and a mass ratio inversion leading to orbital expansion. As a result, ECS-HMXBs have predominantly wide orbits ($\sim$500--2500\,R$_\odot$) and evolved donors. These characteristics, along with their sensitivity to ECS properties, provide us with an intriguing opportunity to probe ECS physics and progenitors through studies of starbursts at different ages. 

The SMC is a promising candidate for a study probing ECS physics. Observations of the SMC have revealed a major burst of star formation 30--60 Myr ago \citep{2001A&A...379..864M, 2004AJ....127.1531H}. Recent observations have found a surprisingly large population of HMXBs with Be donors concentrated in the SMC bar region \citep{2000A&A...359..573H, 2004ApJ...609..133M, 2009ApJ...697.1695A}. 

There are four reasons to prefer the hypothesis that Be HMXBs are primarily formed through ECS creation channels. First, the population of HMXBs in the SMC has been found to have wide orbital separations and low eccentricities, which would require fine tuning of the initial orbital parameters if strong kicks were used \citep{2002ApJ...574..364P}. Second, we note the coincidence of age between the peak of ECS driven HMXB activity, and the observed age of star forming regions of the SMC bar. Third, the spatial distribution of SMC HMXBs is highly peaked in the bar region, with few HMXBs elsewhere in the galaxy. This spatial distribution is highly unlikely to result from high velocity natal kicks occuring between 10 and 40 Myr before the present. Lastly, \citet{2005ApJS..161..118M} have recently argued that Be stars form through spin up due to mass transfer in binaries that experience mass ratio inversion and associated orbital expansion, which fits the evolution pathway found for the ECS HMXBs in our models.

\begin{figure}
		\plotone{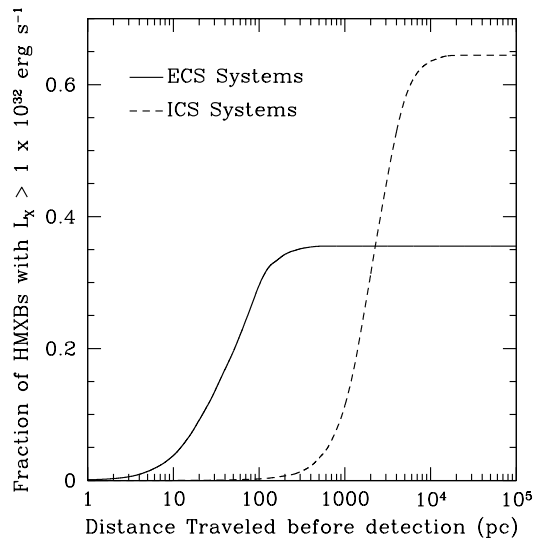}
		\caption{Distance travelled by HMXBs with $L_X>10^{32}$\,erg\,s$^{-1}$ detected between 30-60 Myr after star formation for ECS SN with natal kick velocities of 26.5 km s$^{-1}$. Distances are calculated ignoring any deceleration effects from the underlying galactic gravitational potential.}
		\label{dist}
\end{figure}

In order to estimate the overconcentration of ECS sources in the SMC bar, we select a value of $\sigma_{ECS}$~=~26.5~km~s$^{-1}$, and examine the number of HMXB with L$_X$ $>$ 1 x 10$^{32}$\,erg\,s$^{-1}$ between 30 and 60 Myr after star formation, taking snapshots of the systems position every 0.1 Myr. In Figure~\ref{dist}, we plot the cumulative distribution function of three-dimensional distances traveled due to the SN natal kick of both ECS and ICC sources, normalized to their total number. We ignore any effects due to the galactic gravitational potential on the motion of HMXB sources. We note that while overall there are approximately twice as many ICC sources, only around 15\% of ICC sources are found within 1 kpc of the location in which they were formed. Inside this region, ECS sources dominate by a factor of approximately 4. We note that the deep $Chandra$ observations concentrate on the SMC bar region, while regions further outside are either not targeted by observations or their X-ray luminosity sensitivity limit is much higher. Consequently a significant number of XRBs predicted by the models to have moved to distances beyond $\sim10$ kpc could remain undetected.

Our above findings strongly depend on our choice to set natal kicks for direct collapse BHs to be 10\% as strong as our Maxwellian natal kick distribution. If we set direct collapse natal kicks to be strictly zero, we find a large population of persistent sources in the SMC bar with luminosity $L_X>10^{35}$\,erg\,s$^{-1}$, containing CHeB donors and massive BH accretors, and having orbital periods on the order of thousands of days. As these systems are not only unobserved in the SMC, but throughout all HMXB populations, we enable small direct collapse kicks which split these loosely bound systems. 

Lastly, while the Be star phenomena is not understood from first principles, making it difficult to determine the absolute number of systems with Be-driven luminosities above $L_X>10^{34}$\,erg\,s$^{-1}$, we note that our simulations are consistent with the lack of observed supergiants with spherically driven winds. For each million initial binaries, we find approximately 7.5 such sources with luminosity $L_X>10^{34}$\,erg\,s$^{-1}$ which have travelled less than 1 kpc. Using standard assumptions for initial mass function and star formation rate in the bar region, this is consistent with a Poisson distribution of supergiant systems. 

These results demonstrate the promise of using the ECS process in order to correctly model the great overabundance of Be-HMXB sources in the SMC. Because of the low natal kicks imparted to ECS systems, HMXBs formed through the ECS process are able to replicate the long period of Be-HMXB sources and high concentration of these systems in the SMC bar. While several models exist which allow for low natal kick velocities, we show that the ECS process should be favored due to its ability to create a dominant population of systems at times matching SMC observations. In addition, the ECS mechanism produces HMXBs through the same physical process suspected to produce a Be-HXMB population \citep{2005ApJS..161..118M}. Detailed modeling of the HMXB population in the SMC, taking into account the duration, age and spatial distribution of star formation activity, the motion of the star forming clusters, and the lower metallicity of the SMC is beyond the scope of this {\em paper}, but is currently underway. We anticipate that such studies of the SMC and potentially other nearby galaxies with recent starburst activity will greatly contribute to efforts of understanding SNe and compact object formation.

\acknowledgements
We thank Valia Antoniou, Jay Gallagher, and Andreas Zezas for useful discussions regarding recent observations of starbursts and high-mass X-ray binaries. This work was supported by the NSF CAREER grant AST-0449558 and a Packard Fellowship in Science \& Engineering to VK. Numerical simulations were performed on the HPC cluster {\tt fugu} available to the Theoretical Astrophysics Group at Northwestern University through the NSF MRI grant PHY-0619274 to V.K.

\bibliography{bib1}

\begin{thebibliography}{27}
\expandafter\ifx\csname natexlab\endcsname\relax\def\natexlab#1{#1}\fi

\bibitem[{{Abt}(1983)}]{1983ARA&A..21..343A}
{Abt}, H.~A. 1983, \araa, 21, 343

\bibitem[{{Antoniou} {et~al.}(2009){Antoniou}, {Zezas}, {Hatzidimitriou}, \&
  {McDowell}}]{2009ApJ...697.1695A}
{Antoniou}, V., {Zezas}, A., {Hatzidimitriou}, D., \& {McDowell}, J.~C. 2009,
  \apj, 697, 1695

\bibitem[{{Belczynski} {et~al.}(2008){Belczynski}, {Kalogera}, {Rasio}, {Taam},
  {Zezas}, {Bulik}, {Maccarone}, \& {Ivanova}}]{2008ApJS..174..223B}
{Belczynski}, K., {Kalogera}, V., {Rasio}, F.~A., {Taam}, R.~E., {Zezas}, A.,
  {Bulik}, T., {Maccarone}, T.~J., \& {Ivanova}, N. 2008, \apjs, 174, 223

\bibitem[{{Dessart} {et~al.}(2006){Dessart}, {Burrows}, {Ott}, {Livne}, {Yoon},
  \& {Langer}}]{2006ApJ...644.1063D}
{Dessart}, L., {Burrows}, A., {Ott}, C.~D., {Livne}, E., {Yoon}, S.-C., \&
  {Langer}, N. 2006, \apj, 644, 1063

\bibitem[{{Fragos} {et~al.}(2009){Fragos}, {Willems}, {Kalogera}, {Ivanova},
  {Rockefeller}, {Fryer}, \& {Young}}]{2009ApJ...697.1057F}
{Fragos}, T., {Willems}, B., {Kalogera}, V., {Ivanova}, N., {Rockefeller}, G.,
  {Fryer}, C.~L., \& {Young}, P.~A. 2009, \apj, 697, 1057

\bibitem[{{Haberl} \& {Sasaki}(2000)}]{2000A&A...359..573H}
{Haberl}, F., \& {Sasaki}, M. 2000, \aap, 359, 573

\bibitem[{{Harris} \& {Zaritsky}(2004)}]{2004AJ....127.1531H}
{Harris}, J., \& {Zaritsky}, D. 2004, \aj, 127, 1531

\bibitem[{{Heggie}(1975)}]{1975MNRAS.173..729H}
{Heggie}, D.~C. 1975, \mnras, 173, 729

\bibitem[{{Hobbs} {et~al.}(2005){Hobbs}, {Lorimer}, {Lyne}, \&
  {Kramer}}]{2005MNRAS.360..974H}
{Hobbs}, G., {Lorimer}, D.~R., {Lyne}, A.~G., \& {Kramer}, M. 2005, \mnras,
  360, 974

\bibitem[{{Hurley} {et~al.}(2000){Hurley}, {Pols}, \&
  {Tout}}]{2000MNRAS.315..543H}
{Hurley}, J.~R., {Pols}, O.~R., \& {Tout}, C.~A. 2000, \mnras, 315, 543

\bibitem[{{Ivanova} {et~al.}(2007){Ivanova}, {Heinke}, {Rasio}, {Belczynski},
  \& {Fregeau}}]{2007arXiv0706.4096I}
{Ivanova}, N., {Heinke}, C., {Rasio}, F.~A., {Belczynski}, K., \& {Fregeau}, J.
  2007, ArXiv e-prints, 706

\bibitem[{{Ivanova} \& {Taam}(2004)}]{2004ApJ...601.1058I}
{Ivanova}, N., \& {Taam}, R.~E. 2004, \apj, 601, 1058

\bibitem[{{Kitaura} {et~al.}(2006){Kitaura}, {Janka}, \&
  {Hillebrandt}}]{2006A&A...450..345K}
{Kitaura}, F.~S., {Janka}, H.-T., \& {Hillebrandt}, W. 2006, \aap, 450, 345

\bibitem[{{Majid} {et~al.}(2004){Majid}, {Lamb}, \&
  {Macomb}}]{2004ApJ...609..133M}
{Majid}, W.~A., {Lamb}, R.~C., \& {Macomb}, D.~J. 2004, \apj, 609, 133

\bibitem[{{Maragoudaki} {et~al.}(2001){Maragoudaki}, {Kontizas}, {Morgan},
  {Kontizas}, {Dapergolas}, \& {Livanou}}]{2001A&A...379..864M}
{Maragoudaki}, F., {Kontizas}, M., {Morgan}, D.~H., {Kontizas}, E.,
  {Dapergolas}, A., \& {Livanou}, E. 2001, \aap, 379, 864

\bibitem[{{McSwain} \& {Gies}(2005)}]{2005ApJS..161..118M}
{McSwain}, M.~V., \& {Gies}, D.~R. 2005, \apjs, 161, 118

\bibitem[{{Miyaji} {et~al.}(1980){Miyaji}, {Nomoto}, {Yokoi}, \&
  {Sugimoto}}]{1980PASJ...32..303M}
{Miyaji}, S., {Nomoto}, K., {Yokoi}, K., \& {Sugimoto}, D. 1980, \pasj, 32, 303

\bibitem[{{Nomoto}(1984)}]{1984ApJ...277..791N}
{Nomoto}, K. 1984, \apj, 277, 791

\bibitem[{{Nomoto}(1987)}]{1987ApJ...322..206N}
---. 1987, \apj, 322, 206

\bibitem[{{Pfahl} {et~al.}(2002){Pfahl}, {Rappaport}, {Podsiadlowski}, \&
  {Spruit}}]{2002ApJ...574..364P}
{Pfahl}, E., {Rappaport}, S., {Podsiadlowski}, P., \& {Spruit}, H. 2002, \apj,
  574, 364

\bibitem[{{Podsiadlowski} {et~al.}(2004){Podsiadlowski}, {Langer},
  {Poelarends}, {Rappaport}, {Heger}, \& {Pfahl}}]{2004ApJ...612.1044P}
{Podsiadlowski}, P., {Langer}, N., {Poelarends}, A.~J.~T., {Rappaport}, S.,
  {Heger}, A., \& {Pfahl}, E. 2004, \apj, 612, 1044

\bibitem[{{Poelarends} {et~al.}(2008){Poelarends}, {Herwig}, {Langer}, \&
  {Heger}}]{2008ApJ...675..614P}
{Poelarends}, A.~J.~T., {Herwig}, F., {Langer}, N., \& {Heger}, A. 2008, \apj,
  675, 614

\bibitem[{{Siess}(2007)}]{2007A&A...476..893S}
{Siess}, L. 2007, \aap, 476, 893

\bibitem[{{Taam} \& {Sandquist}(2000)}]{2000ARA&A..38..113T}
{Taam}, R.~E., \& {Sandquist}, E.~L. 2000, \araa, 38, 113

\bibitem[{{van den Heuvel}(2004)}]{2004ESASP.552..185V}
{van den Heuvel}, E.~P.~J. 2004, in ESA Special Publication, Vol. 552, 5th
  INTEGRAL Workshop on the INTEGRAL Universe, ed. V.~{Schoenfelder},
  G.~{Lichti}, \& C.~{Winkler}, 185--+

\bibitem[{{van den Heuvel}(2007)}]{2007AIPC..924..598V}
{van den Heuvel}, E.~P.~J. 2007, in American Institute of Physics Conference
  Series, Vol. 924, American Institute of Physics Conference Series, 598--606

\bibitem[{{Webbink}(1984)}]{1984ApJ...277..355W}
{Webbink}, R.~F. 1984, \apj, 277, 355

\end{thebibliography}

\end{document}